\documentclass[amsmath,amssymb,nofootinbib]{revtex4}
\usepackage{graphicx}
\usepackage{dcolumn}
\usepackage{color}
\usepackage{bm}
\usepackage{amsmath,amssymb,graphicx}
\usepackage{epsfig}
\usepackage{enumerate}
\usepackage{array}
\usepackage{multirow}

\begin{document}
\title{Dynamical symmetries of the anisotropic oscillator}

\author{Akash Sinha\footnote{E-mail: s23ph09005@iitbbs.ac.in}$^1$,  Aritra Ghosh\footnote{E-mail: ag34@iitbbs.ac.in, ~aritraghosh500@gmail.com}$^1$, and Bijan Bagchi\footnote{E-mail: bbagchi123@gmail.com}$^2$}

\vspace{2mm}

\affiliation{$^{1}$School of Basic Sciences, Indian Institute of Technology Bhubaneswar, Jatni, Khurda, Odisha 752050, India\\
$^{2}$Shiv Nadar University, Physics Dept., Gautam Buddha Nagar, Uttar Pradesh 203207, India}

\vskip-2.8cm
\date{\today}
\vskip-0.9cm

\begin{abstract}
It is well known that the Hamiltonian of an $n$-dimensional isotropic oscillator admits an $SU(n)$ symmetry, making the system maximally superintegrable. However, the dynamical symmetries of the anisotropic oscillator are much more subtle. We introduce a novel set of canonical transformations that map an $n$-dimensional anisotropic oscillator to the corresponding isotropic problem. Consequently, the anisotropic oscillator is found to possess the same number of conserved quantities as the isotropic oscillator, making it maximally superintegrable too (commensurate case). The first integrals are explicitly calculated in the case of a two-dimensional anisotropic oscillator and remarkably, they admit closed-form expressions. 
 \end{abstract}
 
 \maketitle

\vspace{7mm}

\textbf{Note:} This version of the article reflects, to the extent deemed reasonable by the authors, post-publication corrections of typographical and minor technical errors, as well as important clarifying remarks in Appendix \ref{add}. The contents and results otherwise match with those of the published version [Phys. Scr. \textbf{98}, 095253 (2023)].

\section{Introduction}\label{introsec}
The harmonic oscillator plays a pivotal role in our understanding of various physical phenomena, starting from classical dynamical systems to those in quantum field theories. It admits a Hamiltonian description as given by the equations of motion:
\begin{equation}
\dot{q}_j = \frac{\partial H}{\partial p_j}, \hspace{5mm} \dot{p}_j = - \frac{\partial H}{\partial q_j}, \quad\quad j \in \{1, 2, \cdots ,n\},
\end{equation} where $H$ is quadratic in $n$ pairs of $(q_j,p_j)$, in the phase space defined for $n$ degrees of freedom. 
For the simplest case of the isotropic oscillator, the Hamiltonian reads 
\begin{equation}\label{Hisopq}
H = \frac{1}{2} \sum_{j = 1}^n \big(p_j^2 +\omega_0^2 q_j^2 \big),
\end{equation}
which is a linear sum, i.e. $H = \sum_{j=1}^n H_j$, with $H_j$ being the component Hamiltonian of the $j$th oscillator having coordinate $q_j$, momentum $p_j$, and frequency (independent of $j$) $\omega_0$. Thus, the system executes simple harmonic motion on each $q_j-p_j$ plane in the phase space with the same frequency $\omega_0$. As is well known, the system is superintegrable \cite{evans}, meaning that it possesses more functionally independent constants of motion than what is required for Liouville-Arnold integrability \cite{arnoldcm}\footnote{For a general classification of classical superintegrable systems with the help of ladder operators, see \cite{marquette}.}. The symmetry group of the Hamiltonian is $SU(n)$ \cite{ANISO,C}, and the conserved quantities follow the $su(n)$ Lie algebra, with respect to the Poisson bracket. It means that all the conserved quantities are not in involution, i.e. they do not admit mutually commuting Poisson brackets. This is certainly expected of any superintegrable system\footnote{It should be noted that any integrable Hamiltonian system is superintegrable in the neighborhood of any regular point in the classical phase space (see for example \cite{supTsi}).}. As a matter of fact, the isotropic oscillator is maximally superintegrable, meaning that it possesses the maximum number of independent integrals of motion that a system with a $2n$-dimensional phase space can admit. 

\vspace{2mm}

For concreteness, we focus on the two-dimensional ($n=2$) case which corresponds to a four-dimensional phase space. Then, the Hamiltonian is 
\begin{equation}
H = \frac{p_1^2 + p_2^2}{2} + \frac{\omega_0^2(q_1^2 + q_2^2)}{2} \rightarrow \omega_0 \Bigg(  \frac{p_1^2 + p_2^2}{2} + \frac{q_1^2 + q_2^2}{2} \Bigg),
\end{equation} where we have performed a (canonical) rescaling: $q_j \rightarrow q_j/\sqrt{\omega_0}$ and $p_j \rightarrow \sqrt{\omega_0} p_j$ for $j = 1,2$. The Hamiltonian admits an $SU(2)$ symmetry and the corresponding conserved quantities are $I_1 = p_1p_2 + q_1 q_2$, $I_2 = q_2 p_1 - q_1 p_2$, and $I_3 = (p_1^2 + q_1^2) - (p_2^2 + q_2^2)$. The quantities $I_1$ and $I_2$ are, respectively, the Fradkin tensor and the angular momentum on the $q_1-q_2$ plane (with a negative sign), while if we take into account the fact that the Hamiltonian is conserved because it Poisson-commutes with itself (we denote it by $I_0$), then $I_3$ gives rise to the following two conserved quantities: $I^a_3 = p_1^2 + q_1^2$ and $I^b_3 = p_2^2 + q_2^2$, which correspond to the conserved energies in each individual degree of freedom. Notice that since the phase space is four-dimensional, a maximally superintegrable system shall possess only three conserved quantities that are functionally independent. It is easy to check that the first integrals $I_1$, $I_2$, $I^a_3$, and $I^b_3$ are not all independent from one another. The same arguments go through for the general case with a $2n$-dimensional phase space.  

\vspace{2mm}

Although the isotropic oscillator does enjoy such an extended set of symmetries and therefore, conserved quantities \cite{vinet}, the situation gets much more complicated for the anisotropic oscillator. The anisotropic oscillator is one, for which the Hamiltonian is $H = \sum_{j=1}^n H_j$, where $H_j$ is the Hamiltonian of the $j$th harmonic oscillator with coordinate $q_j$, momentum $p_j$, and frequency $\omega_j$. Moreover, it should be remarked that we make no assumptions on the commensurability of the frequencies (see Appendix \ref{add} for clarification). We want to ask: what are the symmetries and conserved quantities of the anisotropic oscillator? Clearly, the anisotropic oscillator is not a central force problem and therefore, does not admit conservation of angular momentum. However, as we shall later show, the system admits a hidden $SU(n)$ symmetry which is not apparent when we apply the $(q_j,p_j)$ variables. 

\vspace{2mm}

The question of finding the first integrals of the anisotropic oscillator was investigated several years ago in \cite{ANISO}, where it was reported that the (commensurate) anisotropic oscillator possesses the same number of symmetries as the isotropic oscillator, but holding on rather restricted regions in the phase space. In this paper, we report further progress by introducing novel canonical transformations which map the anisotropic oscillator to the corresponding isotropic one. If such a canonical transformation can be found, then the dynamical symmetries of the transformed anisotropic oscillator could be related to the $SU(n)$ group. Further, by inverting the transformations, one can determine the conserved quantities of the anisotropic oscillator, from the knowledge of those of its isotropic counterpart. In this way, the anisotropic oscillator can be seen to possess a hidden $SU(n)$ symmetry that can be uncovered by a sequence of canonical transformations, which we will systematically describe in the following section. Since Poisson bracket relations are preserved under canonical transformations, the first integrals, even when expressed in the original set of oscillator variables also admit the same $su(n)$ Lie algebra, putting the anisotropic oscillator on a similar footing as the isotropic oscillator. The isotropic oscillator appears only as a special case of the anisotropic oscillator by fixing $\omega_1 = \omega_2 = \cdots = \omega_n$. 

\vspace{2mm}

The contents of this paper are as follows\footnote{We do not use the summation convention for repeated indices. For instance, $a_j b_j = f_j$ is not being summed over all values of $j$. In order to denote a sum, we shall write $\sum_{j=1}^n f_j = \sum_{j=1}^n a_j b_j$.}: In Section \ref{DSsec}, we discuss the dynamical symmetries of the anisotropic oscillator by considering a sequence of canonical transformations that map it to the isotropic oscillator. The transformations are derived in the Section \ref{cansec}, while the associated generating functions are presented in Section \ref{genfuncsec}. In Section \ref{consec}, we explicitly compute the first integrals of the anisotropic oscillator, where the first integrals appear as appropriate generalizations of the ones appearing for the isotropic oscillator. Some concluding remarks are presented at the end in Section \ref{dsec}. 

\section{Dynamical symmetries}\label{DSsec}
The basic sequence of steps for finding the aforementioned transformation is as follows. Let us consider an $n$-dimensional isotropic oscillator for which the Hamiltonian is given by Eq. (\ref{Hisopq}). With respect to the scaling $q_j \rightarrow q_j/\sqrt{\omega_0}$ and $p_j \rightarrow \sqrt{\omega_0} p_j$, $j = 1,2,\cdots n$, the Hamiltonian can be converted to the form:
\begin{equation}
H = \frac{\omega_0}{2} \sum_{j=1}^n (p_j^2 + q_j^2).
\end{equation}
Defining a new pair of variables that still preserve the canonical structure:
\begin{equation}\label{XP}
X_j = \frac{q_j - i p_j}{\sqrt{2}}, \hspace{5mm} P_j = \frac{p_j - i q_j}{\sqrt{2}},
\end{equation} the Hamiltonian reads
\begin{equation}\label{HPX}
H = i \omega_0 \sum_{j=1}^n P_j X_j.
\end{equation}
Observe that the transformations given in Eq. (\ref{XP}) have a remarkable property, namely $P_j = - iX_j^* $, where `$*$' denotes complex conjugation. A consequence of this is that the Hamiltonian of Eq. (\ref{HPX}) is invariant under $SU(n)$ transformations and therefore, the underlying generators of the infinitesimal transformations correspond to the conserved quantities. 

\vspace{2mm}

Let us now turn to the anisotropic oscillator in $n$ dimensions. Performing the canonical rescaling transformation, i.e. $q_j \rightarrow q_j/\sqrt{\omega_j}$ and $p_j \rightarrow \sqrt{\omega_j} p_j$, $j = 1,2,\cdots n$, followed by the ones in Eq. (\ref{XP}), one easily finds the following representation:
\begin{equation}\label{HPXaniso}
H = i \omega_0  \sum_{j=1}^n \Omega_j P_j X_j, \hspace{5mm} \Omega_j = \frac{\omega_j}{\omega_0}. 
\end{equation}
However, the Hamiltonian is no longer symmetric under $SU(n)$ transformations because of the appearance of the dimensionless coefficient factors\footnote{We may choose $\omega_0:=\frac{1}{n} \sum_{j=1}^n \omega_j $. For the isotropic case, this yields $\Omega_j=\frac{\omega_j}{\omega_0} = 1$, $\forall j$.} $\{\Omega_j\}$, which are different from each other. Nonetheless, notice that Eq. (\ref{HPXaniso}) admits a $U(1)$ symmetry on each $X_j-P_j$ plane, implying the existence of the underlying symmetry group $U(1) \oplus U(1) \oplus \cdots \oplus U(1)$ ($n$ times) in play. The symmetry group is obviously smaller than $U(n)$ or $SU(n)$, and the conserved quantities are $I_j =\omega_j( p_j^2 + q_j^2)$ for $j \in \{1,2, \cdots, n\}$, corresponding to the mechanical energies associated with motion in each individual spatial direction. 

\subsection{Canonical transformations between anisotropic and isotropic oscillators}\label{cansec}
We next proceed to show that there are several hidden conserved quantities that can be uncovered by performing another set of canonical transformations. To this end, our task would be to map the Hamiltonian of the anisotropic oscillator given in Eq. (\ref{HPXaniso}), to the isotropic one given in Eq. (\ref{HPX}). We therefore look for a new set of variables $\mathcal{X}_j = \mathcal{X}_j(P_j,X_j)$ and $\mathcal{P}_j = \mathcal{P}_j(P_j,X_j)$, such that the following conditions are met: 
\begin{enumerate}
\item The new variables ought to satisfy $\Omega_j P_j X_j = \mathcal{P}_j \mathcal{X}_j$.
\item The transformation should be canonical, i.e. $ \{\mathcal{X}_j,\mathcal{P}_k\} = \delta_{jk}$ and $ \{\mathcal{X}_j,\mathcal{X}_k\} = \{\mathcal{P}_j,\mathcal{P}_k\} = 0$, where $\{\cdot,\cdot\}$ is the Poisson bracket evaluated in the $X_j,P_j$ basis. 
\item For each $j \in \{1,2, \cdots, n\}$, one has to have $\mathcal{P}_j = - i\mathcal{X}_j^* $.
\end{enumerate}

In the following discussion, we suppress the index $j$ and focus on two general functions $\mathcal{X}=\mathcal{X}(X,P)$ and $\mathcal{P}=\mathcal{P}(X,P)$, where we assume that $\mathcal{P} \neq 0$. Now, the first condition implies 
\begin{eqnarray}\label{calx}
\mathcal{X}(X,P)=\frac{\Omega PX}{\mathcal{P}(X,P)}.
\end{eqnarray} Next, the second condition gives
\begin{equation}
\frac{\partial \mathcal{X}}{\partial X} \frac{\partial \mathcal{P}}{\partial P}  - \frac{\partial \mathcal{P}}{\partial X} \frac{\partial \mathcal{X}}{\partial P} = 1,
\end{equation}
which, when Eq. (\ref{calx}) is used, furnishes the partial differential equation:
\begin{eqnarray}
P \frac{\partial \mathcal{P}}{\partial P}-X\frac{\partial \mathcal{P}}{\partial X}&=& \frac{\mathcal{P}}{\Omega}.
\end{eqnarray}
This differential equation can be solved through separation of variables. Let us make the ansatz $\mathcal{P}(X,P)=\alpha(X)\beta(P)$, where $\alpha(X), \beta(P) \neq 0$ since $\mathcal{P} \neq 0$. Then, it follows that 
\begin{eqnarray}
\Omega \left[\frac{P}{\beta} \frac{\partial \beta}{\partial P}-\frac{X}{\alpha}\frac{\partial \alpha}{\partial X}\right] =1. 
\end{eqnarray}
Introducing a separation constant $k$ yields two ordinary differential equations:
\begin{equation}
\Omega \frac{P}{\beta} \frac{d \beta}{d P}=k+\frac{1}{2}, \hspace{5mm} \Omega \frac{X}{\alpha}\frac{d \alpha}{d X}=k-\frac{1}{2}.
\end{equation}
These can be solved to obtain 
\begin{eqnarray}
\beta(P)=\beta_0 P^{\frac{1}{\Omega}\left(k+\frac{1}{2}\right)},\quad\quad \alpha(X)=\alpha_0 X^{\frac{1}{\Omega}\left(k-\frac{1}{2}\right)}.
\end{eqnarray}
We thus have the following closed-form expressions:
\begin{eqnarray}
\mathcal{P}(X,P)&=&AX^{\frac{1}{\Omega}\left(k-\frac{1}{2}\right)}P^{\frac{1}{\Omega}\left(k+\frac{1}{2}\right)},\\
\mathcal{X}(X,P)&=&\frac{\Omega}{A} X^{1-\frac{1}{\Omega}\left(k-\frac{1}{2}\right)}P^{1-\frac{1}{\Omega}\left(k+\frac{1}{2}\right)},
\end{eqnarray} where $A$ is a constant which can be determined from the third condition that is to be accounted for. Some straightforward calculations lead to 
\begin{eqnarray}
|A|^2=\Omega,\quad\quad k=\Omega/2.
\end{eqnarray} Restoring now the index $j$, we have the results:
\begin{eqnarray}
\mathcal{X}_j(X_j,P_j)&=&\sqrt{\Omega_j}X_j^{\frac{1}{2}\left(1+\frac{1}{\Omega_j}\right)}P_j^{\frac{1}{2}\left(1-\frac{1}{\Omega_j}\right)}, \label{1x} \\
 \mathcal{P}_j(X_j,P_j)&=&\sqrt{\Omega_j}X_j^{\frac{1}{2}\left(1-\frac{1}{\Omega_j}\right)}P_j^{\frac{1}{2}\left(1+\frac{1}{\Omega_j}\right)}. \label{1p}
\end{eqnarray}
The above forms are new and of interest, in that they specify the canonical transformations that map the Hamiltonian of the anisotropic oscillator to that of the isotropic one. Thus, in these new canonical variables, the anisotropic oscillator admits an $SU(n)$ symmetry. 

\subsection{Generating functions}\label{genfuncsec}
Our aim now would be to obtain the generating functions corresponding to the canonical transformations $(X_j,P_j) \rightarrow (\mathcal{X}_j,\mathcal{P}_j)$. Since the transformation is canonical, we have \cite{goldsteincm,bagchicm}
\begin{equation}\label{canform}
\sum_{j=1}^n P_j \dot{X}_j - H = \sum_{j=1}^n \mathcal{P}_j \dot{\mathcal{X}}_j - K + \frac{dF}{dt},
\end{equation} where $H$ is the Hamiltonian given by Eq. (\ref{HPXaniso}),  $K$ is the Hamiltonian of the transformed system, i.e. the isotropic oscillator, and $F$ is the generating function. Explicitly, one has\footnote{We set $\omega_0 = 1$ for the sake of brevity in our discussion of generating functions.} $K = i  \sum_{j=1}^n  \mathcal{P}_j \mathcal{X}_j$. Notice that the left-hand side of Eq. (\ref{canform}) is just $\theta_H/dt$, where $\theta_H = \sum_{j=1}^n P_j dX_j - H dt$ is the Poincar\'e-Cartan one-form on the extended phase space $\mathcal{M} \times \mathbb{R}$, with $\mathcal{M}$ being the original phase space, which is a symplectic manifold. On $\mathcal{M}$, the local coordinates (namely, the Darboux coordinates) are $(X_j,P_j)$, while $t \in \mathbb{R}$ is a global coordinate. Now, the right-hand side leads to the transformed Poincar\'e-Cartan form \cite{arnoldcm}, say $\theta'_K$, such that $\int \theta_H = \int \theta'_K$ between two fixed end points, where the total differential $dF$ from the right-hand side integrates to zero.

\vspace{2mm}

Taking $F$ to be a function of the original coordinates and transformed coordinates, i.e. $F = F(X_j, \mathcal{X}_j)$, it follows from Eq. (\ref{canform}) that
\begin{equation}\label{abcd}
P_j = \frac{\partial F(X_j, \mathcal{X}_j)}{\partial X_j}, \hspace{5mm} \mathcal{P}_j = -\frac{\partial F(X_j, \mathcal{X}_j)}{\partial \mathcal{X}_j}.
\end{equation}
It is trivial to verify that the following choice:
\begin{equation}\label{F1def}
F(X_j,\mathcal{X}_j)=\frac{1}{2} \sum_{j=1}^n (1-\Omega_j)\Omega_j^{\frac{\Omega_j}{1-\Omega_j}}X_j^{\frac{2}{1-\Omega_j}}\mathcal{X}_j^{-\frac{2\Omega_j}{1-\Omega_j}},
\end{equation}
is consistent with Eq. (\ref{abcd}) and is therefore a generator of the canonical transformation. It is of course, possible to define other forms of generating functions that involve the momentum variables. From Eq. (\ref{canform}), it is clear that they would be related by Legendre transforms because the symplectic potential\footnote{For a mechanical system such as an oscillator, we may consider the phase space to be a cotangent bundle $\mathcal{M} = T^*Q$, where $Q$ is the configuration space. In that case, the phase space is an exact symplectic manifold, i.e. the symplectic form reads $\omega_s = d\theta_s$, where $\theta_s$ is the tautological one-form on the vector bundle $T^*Q$. In a general case however, a symplectic manifold need not be exact, i.e. the symplectic form may not be exact, although it is always closed by definition.} $\theta_s$ (whose exterior derivative gives the symplectic form on $\mathcal{M}$, i.e. $\omega_s = d\theta_s$) gives rise to the same $\omega_s$ up to Legendre transforms, which are merely a subclass of canonical transformations (symplectomorphisms) preserving $\omega_s$. For completeness, we have listed in Table I, four classes of generating functions and the Legendre transforms connecting them, in which we denote the one appearing in Eq. (\ref{F1def}) as $F_1$. 

\begin{table*}[t]
\caption{Generating functions.\\}
\centering
\begin{tabular}{| c | c | c | c |}
\hline
{\bf Generating function} & {\bf Explicit expression} &  {\bf Total differential}  &  {\bf Relation with $F_1(X_j,\mathcal{X}_j)$}    \\ \hline
$F_1(X_j,\mathcal{X}_j)$    &   $\frac{1}{2} \sum_{j=1}^n(1-\Omega_j)\Omega_j^{\frac{\Omega_j}{1-\Omega_j}}X_j^{\frac{2}{1-\Omega_j}}\mathcal{X}_j^{-\frac{2\Omega_j}{1-\Omega_j}}$  &  $dF_1 = \sum_{j=1}^n (P_j dX_j - \mathcal{P}_j d \mathcal{X}_j)$      &  $-$   \\
$F_2(X_j,\mathcal{P}_j)$      &       $ \frac{1}{2} \sum_{j=1}^n(1+\Omega_j)\Omega_j^{-\frac{\Omega_j}{1+\Omega_j}}X_j^{\frac{2}{1+\Omega_j}}\mathcal{P}_j^{\frac{2\Omega_j}{1+\Omega_j}}  $   &  $dF_2 =  \sum_{j=1}^n (P_j dX_j + \mathcal{X}_j d \mathcal{P}_j)$       &  $F_2 = F_1 + \sum_{j=1}^n \mathcal{X}_j \mathcal{P}_j$  \\ 
 $F_3(P_j,\mathcal{X}_j)$      &       $ -\frac{1}{2} \sum_{j=1}^n(1+\Omega_j)\Omega_j^{-\frac{\Omega_j}{1+\Omega_j}}P_j^{\frac{2}{1+\Omega_j}}\mathcal{X}_j^{\frac{2\Omega_j}{1+\Omega_j}}  $   &   $dF_3 =  \sum_{j=1}^n (-X_j dP_j - \mathcal{P}_j d\mathcal{X}_j )$      &  $F_3 = F_1 - \sum_{j=1}^n X_j P_j$  \\ 
$F_4(P_j,\mathcal{P}_j)$      &       $ -\frac{1}{2} \sum_{j=1}^n (1-\Omega_j)\Omega_j^{\frac{\Omega_j}{1-\Omega_j}}P_j^{\frac{2}{1-\Omega_j}}\mathcal{P}_j^{-\frac{2\Omega_j}{1-\Omega_j}}  $      &  $dF_4 =  \sum_{j=1}^n (-X_j dP_j +  \mathcal{X}_j d\mathcal{P}_j) $    &  $F_4 = F_1 + \sum_{j=1}^n (\mathcal{X}_j \mathcal{P}_j - X_j P_j)$  \\ \hline
\end{tabular}
\end{table*}

\section{Conserved quantities}\label{consec}
We now determine the conserved quantities, or first integrals associated with the anisotropic oscillator. For simplicity, we pick $n=2$ and the symmetry group is $SU(2)$.
We write the conserved quantities in the form: $\mathcal{I}_{\mu}=i(\sigma_{\mu})_{jk}\mathcal{P}_j\mathcal{X}_k$, with $\mu=0,1,2,3$. Here, $\sigma_0 = I_2$ (identity) and $\sigma_1$,  $\sigma_2$, $\sigma_3$ are the Pauli matrices\footnote{The conserved quantities arising from $SU(2)$ (three generators) correspond to the indices $\mu = 1,2,3;$ whereas the quantity $\mathcal{I}_0$ can be understood to originate from the extended symmetry group $U(2)$. We are slightly abusive with our notation by indicating that the symmetry group is $SU(n)$ in general. Actually the system has a $U(n)$ symmetry such that the non-trivial conserved quantities originating from $SU(n)$, together with the Hamiltonian of the oscillator, i.e. $\mathcal{I}_0$, form the set of conserved quantities. This is discussed in Section \ref{dsec}.}. A straightforward generalization can be done for a higher number of canonical variables. 

\vspace{2mm}

 We first consider the quantity $\mathcal{I}_0=i(\sigma_0)_{jk}\mathcal{P}_j \mathcal{X}_k$. It takes the form:
\begin{eqnarray}
\mathcal{I}_0&=&i(\mathcal{P}_1\mathcal{X}_1+\mathcal{P}_2\mathcal{X}_2) \nonumber \\
&=&i(\Omega_1P_1X_1+\Omega_2P_2X_2), 
\end{eqnarray}
which is just the total energy of the system (in units of $\omega_0$). The quantity $\mathcal{I}_3$ can be obtained similarly and is analogous to $I_3$ discussed for the isotropic case. The analogues of angular momentum and Fradkin tensor, correspond, respectively, to $\mathcal{I}_2=\mathcal{P}_1\mathcal{X}_2-\mathcal{P}_2\mathcal{X}_1$ and $\mathcal{I}_1=i(\mathcal{P}_1\mathcal{X}_2+\mathcal{P}_2\mathcal{X}_1)$. Specifically for $\mathcal{I}_2$, we find 
\begin{eqnarray}
\mathcal{I}_2 &=& \left({\Omega_1P_1X_1\Omega_2P_2X_2}\right)^{\frac{1}{2}}  \left[(P_1/X_1)^{\frac{1}{2\Omega_1}}(X_2/P_2)^{\frac{1}{2\Omega_2}}-(X_1/P_1)^{\frac{1}{2\Omega_1}}(P_2/X_2)^{\frac{1}{2\Omega_2}}\right],  \label{z2}
\end{eqnarray} which clearly goes to $I_2=P_1X_2-P_2X_1$ if $\omega_1 = \omega_2 = \omega_0$ or $\Omega_1 = \Omega_2 = 1$ (isotropic limit). We consider two possibilities below and obtain the first integrals in each case. 

\subsection{Nearly equal frequencies: $\Omega_1 \approx \Omega_2$}
One can consider a special case where the two frequencies slightly differ from each other. To enforce that, we put $\omega_1 = (1 + \epsilon)\omega_0$ and $\omega_2 = (1 - \epsilon)\omega_0$, for $0 < \epsilon << 1$. This means $\Omega_1 - \Omega_2 = 2 \epsilon$ is an infinitesimally small number. In other words, the system departs infinitesimally from its isotropic counterpart and therefore, we expect the conserved quantities to be the same as those of the isotropic case up to perturbative corrections in powers of $\epsilon$ (the deformation parameter). It turns out that this is indeed the case.

\vspace{2mm}

The conserved quantities in terms of the original oscillator variables $(q_1,q_2,p_1,p_2)$ for small $\epsilon$ read
\begin{eqnarray}
\mathcal{I}_0^{\epsilon}&=&\frac{1+\epsilon}{2}(p_1^2+q_1^2)+\frac{1-\epsilon}{2}(p_2^2+q_2^2), \\
\mathcal{I}_1^{\epsilon}&=&(p_1p_2+q_1q_2)-\epsilon(q_2p_1-q_1p_2)\left[(\theta_1+\theta_2)+\frac{\pi}{2}\right]+\mathcal{O}(\epsilon^2),\\
\mathcal{I}_2^{\epsilon}&=&(q_2p_1-q_1p_2)+\epsilon(p_1p_2+q_1q_2)\left[(\theta_1+\theta_2)+\frac{\pi}{2}\right]+\mathcal{O}(\epsilon^2),\\
\mathcal{I}_3^{\epsilon}&=&\frac{1+\epsilon}{2}(p_1^2+q_1^2)-\frac{1-\epsilon}{2}(p_2^2+q_2^2),
\end{eqnarray} where $\theta_j = -\arctan{\left(\frac{p_j}{q_j}\right)}$, and we have used the fact that $X_j=r_j\exp{(i\theta_j)}$ and $P_j=-iX_j^*$, to write
\begin{eqnarray}
\frac{X_j}{P_j}=i\exp{(2i\theta_j)}=\exp{\left[i\left(\frac{\pi}{2}+2\theta_j\right)\right]}.
\end{eqnarray}
We have put the superscript `$\epsilon$' in labeling the conserved quantities to remind the reader that they are obtained in the limit $\Omega_1 \approx \Omega_2$. It is clear that $\mathcal{I}^\epsilon_\mu \approx I_\mu + \cdots$, where $I_\mu$ are the corresponding quantities of the isotropic oscillator, and `$\cdots$' denotes corrections in powers of $\epsilon$ marking the departure from genuine isotropic behavior.

\subsection{General $\Omega_1$ and $\Omega_2$}
We now find the expressions for the first integrals for general values of $\Omega_1$ and $\Omega_2$. We note that $P_1X_1P_2X_2=-(p_1^2+q_1^2)(p_2^2+q_2^2)/4$, and this enables us to obtain the following expressions for $\mathcal{I}_0$, $\mathcal{I}_1$, $\mathcal{I}_2$, and $\mathcal{I}_3$ in terms of the original oscillator variables $(q_1,q_2,p_1,p_2)$:

\begin{eqnarray}
\mathcal{I}_0&=&\frac{\Omega_1}{2}(p_1^2+q_1^2)+\frac{\Omega_2}{2}(p_2^2+q_2^2),\\
\mathcal{I}_1&=&\sqrt{\Omega_1\Omega_2(p_1^2+q_1^2)(p_2^2+q_2^2)}\cos{\left[\frac{\pi}{4}\left(\frac{1}{\Omega_2}-\frac{1}{\Omega_1}\right)+\left(\frac{\theta_2}{\Omega_2}-\frac{\theta_1}{\Omega_1}\right)\right]},\\
\mathcal{I}_2&=&\sqrt{\Omega_1\Omega_2(p_1^2+q_1^2)(p_2^2+q_2^2)}\sin{\left[\frac{\pi}{4}\left(\frac{1}{\Omega_2}-\frac{1}{\Omega_1}\right)+\left(\frac{\theta_2}{\Omega_2}-\frac{\theta_1}{\Omega_1}\right)\right]},\\
\mathcal{I}_3&=&\frac{\Omega_1}{2}(p_1^2+q_1^2)-\frac{\Omega_2}{2}(p_2^2+q_2^2),
\end{eqnarray}
with $\theta_j=-\arctan{\left(\frac{p_j}{q_j}\right)}$. 

\vspace{2mm}

One can indeed check that the conserved quantities listed above Poisson-commute with the Hamiltonian of the two-dimensional anisotropic oscillator, which is just $\mathcal{I}_0$ (up to a factor of $\omega_0$). It may be verified that the expressions for $\mathcal{I}_1$ and $\mathcal{I}_2$ reduce to the familiar expressions for the Fradkin tensor and angular momentum, respectively, for $\Omega_1 = \Omega_2 = 1$. For instance, let us consider $\mathcal{I}_2$ with $\Omega_1 = \Omega_2 = 1$, giving
\begin{eqnarray}
\mathcal{I}_2&=&\sqrt{(p_1^2+q_1^2)(p_2^2+q_2^2)}\sin (\theta_2 - \theta_1) \nonumber \\
&=&\sqrt{(p_1^2+q_1^2)(p_2^2+q_2^2)} \big( \sin \theta_2 \cos \theta_1 - \sin \theta_1 \cos \theta_2 \big). \label{I2isolim}
\end{eqnarray}
Since by definition, $\tan \theta_1 = -p_1/q_1$ and $\tan \theta_2 = -p_2/q_2$, one has 
\begin{equation}
\cos \theta_1 = \frac{q_1}{\sqrt{p_1^2 + q_1^2}}, \hspace{5mm} \cos \theta_2 = \frac{q_2}{\sqrt{p_2^2 + q_2^2}}, \hspace{5mm} \sin \theta_1 = \frac{-p_1}{\sqrt{p_1^2 + q_1^2}}, \hspace{5mm} \sin \theta_2 = \frac{-p_2}{\sqrt{p_2^2 + q_2^2}}.
\end{equation}
Substituting these into Eq. (\ref{I2isolim}) gives $I_2 = q_2 p_1 - q_1 p_2$, as expected. Similarly, $\mathcal{I}_1$ reduces to $I_1 = p_1 p_2 + q_1 q_2$ for $\Omega_1 = \Omega_2 = 1$. 

\vspace{2mm}

Furthermore, the first integrals $\{\mathcal{I}_k\}$ for $k \in \{ 1, 2, 3\}$, follow the $su(2)$ algebra: 
\begin{equation}\label{su2lie}
\{\mathcal{J}_j,\mathcal{J}_k\} = i \epsilon_{jkl} \mathcal{J}_l,
\end{equation} 
where $\{\cdot,\cdot\}$ is a Poisson bracket evaluated in the $(q_1,q_2,p_1,p_2)$ basis, while $\mathcal{J}_k= - i\mathcal{I}_k/2$. The Casimir (central element) of this algebra is $\mathcal{J} =  \mathcal{I}_0$, i.e. $\{\mathcal{J}_k,\mathcal{J}\} = 0$ for $k \in \{ 1, 2, 3\}$, and $\mathcal{J}$ can in principle, be written as a function of $\{\mathcal{J}_k\}$. 

\vspace{2mm}

The individual mechanical energies in the two spatial directions (in units of $\omega_0$) are obtained as $E_1 = (\mathcal{I}_0 + \mathcal{I}_3)/2$ and $E_2 = (\mathcal{I}_0 - \mathcal{I}_3)/2$. The quantities $\mathcal{I}_1$ and $\mathcal{I}_2$ can be viewed as the generalized Fradkin tensor and generalized angular momentum, respectively, although they bear very little resemblance to $I_1$ and $I_2$ of the isotropic oscillator. Nevertheless, one finds that the anisotropic oscillator has the same number of conserved quantities as the isotropic one, as was also discussed in \cite{ANISO} (see further \cite{an1,an2,an3}), and therefore it is a maximally superintegrable system (see Appendix \ref{add} for clarification). A generalization to a higher number of dimensions can be performed straightforwardly, and we do not discuss the issue here. 

\section{Concluding remarks}\label{dsec}
In this paper, we introduced a set of canonical transformations which map the Hamiltonian of the anisotropic oscillator to that of the isotropic oscillator. Since the isotropic oscillator possesses an $SU(n)$ symmetry, this allowed us to compute the conserved quantities of the anisotropic oscillator. The conserved quantities include the mechanical energies in each individual direction. In addition, the anisotropic oscillator has generalized versions of the Fradkin tensor and the angular momentum as conserved quantities. In two dimensions, it is shown that in the case where the two frequencies $\Omega_1$ and $\Omega_2$ are arbitrarily close, then all the conserved quantities reduce to those of the isotropic oscillator together with small corrections originating from the small difference between the frequencies. 

\vspace{2mm}

One should further remark that although we have argued that the symmetry group associated with the $n$-dimensional oscillator (both isotropic and anisotropic) is $SU(n)$, the system admits a slightly larger symmetry group, which is $U(n)$. This is easy to see from our two-dimensional example wherein, the $SU(2)$ generators obeying the $su(2)$ algebra were just three in number, i.e. $\{\mathcal{I}_1, \mathcal{I}_2, \mathcal{I}_3\}$, but we have an additional constant of motion which is $\mathcal{I}_0$. This fourth first integral can be associated with the extended symmetry group $U(2)$ of which $SU(2)$ is a subgroup. However, since on a four-dimensional phase space, only up to three constants of motion can be functionally independent, we may just consider $\{\mathcal{I}_1, \mathcal{I}_2, \mathcal{I}_3\}$ as the independent first integrals, while $\mathcal{I}_0$, which can in principle be expressed in terms of the others is just the Casimir (central element) of the $su(2)$ Lie algebra given in Eq. (\ref{su2lie}). Thus, it suffices to say that the symmetry group is $SU(n)$ with the conserved quantities obeying an $su(n)$ Lie algebra. Finally, we remark that for larger values of $n$, say for $n = 3$, there are more conserved quantities than can be functionally independent coming from the group $SU(n)$, but nevertheless, the quantity $\mathcal{I}_0$ (just the Hamiltonian) comes from the extended group $U(n)$, and Poisson-commutes with all others. \\

\textbf{Acknowledgements:} We thank the anonymous referees for their valuable comments which have led to a considerable improvement of this article. A.S. would like to acknowledge the financial support from IIT Bhubaneswar in the form of an Institute Research Fellowship. The work of A.G. is supported by Ministry of Education (MoE), Government of India in the form of a Prime Minister's Research Fellowship (ID: 1200454). 

\appendix

\section{Addendum}\label{add}
Some clarifying remarks are in order. The canonical transformations introduced in Eqs. (\ref{1x}) and (\ref{1p}) satisfy the defining relations $\mathcal P_j\mathcal X_j=\Omega_j P_jX_j$ and $\{\mathcal X_j,\mathcal P_k\}=\delta_{jk}$ on domains where the fractional powers appearing in their explicit expressions are differentiable. However, these transformations involve non-integer powers of $X_j$ and $P_j$, and therefore require a choice of branch. As a result, they are not single-valued on the punctured complex plane $(X_j,P_j)\in\mathbb C^2\setminus\{X_jP_j=0\}$ unless the exponents are rational and compatible branch choices are imposed. In particular, for irrational $\Omega_j$, the transformations are intrinsically multi-valued and cannot be extended to globally defined symplectomorphisms of the phase space. Consequently, the associated `hidden' conserved quantities are, in general, multi-valued when viewed as functions on the phase space and become globally well defined only under additional conditions.

\vspace{2mm}

The conserved quantities obtained by pulling back the $U(n)$ generators of the isotropic oscillator via the transformations of Section \ref{cansec} are therefore to be understood in a local or covering-space sense. When expressed in action-angle variables, these quantities involve combinations of the form $\theta_j/\Omega_j$, where each angle $\theta_j$ is defined only modulo $2\pi$. For generic (incommensurate) values of the frequencies, such combinations do not descend to single-valued smooth functions on the phase space. They, however, remain constant along the trajectories once the angles are lifted to $\mathbb{R}$. Global, single-valued additional first integrals are recovered precisely in the commensurate case when the relevant frequency ratios are rational and the angle combinations close after a finite winding. With this clarification, the results of this paper should be interpreted as exhibiting a local\footnote{We thank Ond\v{r}ej Kub\r{u} for kindly pointing this out.} realization of the hidden symmetries of the anisotropic oscillator.

\end{document}